\documentclass[twocolumn, prl, superscriptaddress]{revtex4-1}
\usepackage{amssymb}
\usepackage{amsmath}
\usepackage{epsfig}
\usepackage{graphics, graphicx}
\usepackage{bbold}
\usepackage{psfrag}
\usepackage{mathcomp}
\usepackage{subfigure}
\usepackage{verbatim}
\usepackage{color}
\usepackage[colorlinks,citecolor=blue]{hyperref}
\def\cp#1{\mathbf{#1}}

\begin{document}

\title{Universal tetramer and pentamer in two-dimensional fermionic mixtures}
\author{Ruijin Liu}
\affiliation{Beijing National Laboratory for Condensed Matter Physics, Institute of Physics, Chinese Academy of Sciences, Beijing, 100190, China}
\author{Cheng Peng}
\affiliation{Beijing National Laboratory for Condensed Matter Physics, Institute of Physics, Chinese Academy of Sciences, Beijing, 100190, China}
\affiliation{School of Physical Sciences, University of Chinese Academy of Sciences, Beijing 100049, China}
\author{Xiaoling Cui}
\email{xlcui@iphy.ac.cn}
\affiliation{Beijing National Laboratory for Condensed Matter Physics, Institute of Physics, Chinese Academy of Sciences, Beijing, 100190, China}
\affiliation{Songshan Lake Materials Laboratory, Dongguan, Guangdong 523808, China}

\begin{abstract}
We study the emergence of universal tetramer and pentamer bound states in the two-dimensional $(N+1)$ system, which consists of $N$ identical heavy fermions interacting with a light atom. 
We show that the critical heavy-light mass ratio to support a ($3+1$) tetramer  below the trimer threshold is $3.38$, and to support a ($4+1$) pentamer  below the tetramer threshold is $5.14$. 
While these ground state tetramer and pentamer are both with zero total angular momentum, they exhibit very different  density distributions and correlations in momentum space, due to their distinct angular momentum decompositions in the dimer-fermion frame. These universal bound states can be accessible by a number of Fermi-Fermi mixtures now realized in cold atoms laboratories, which also suggest novel few-body correlations dominant in their corresponding many-body systems.
\end{abstract}
\maketitle

Exactly solvable few-body problems provide a crucial and reliable route for approaching the many-body physics. 
As a well-known example, the formation of two-body bound state plays a fundamental role in driving the BCS-BEC crossover\cite{RMP_review} and the polaron-molecule transition\cite{polaron_review, polaron_review2} of spin-1/2 fermions. A natural follow-up question is whether 
there are ways to go beyond two-body and engineer more fascinating few-body bound states in fermion systems, which may evoke even intriguing many-body physics.

One such efficient way is to consider the fermion components with mass-imbalance, which has become accessible by a number of ultracold Fermi-Fermi mixtures such as $^{40}$K-$^{6}$Li\cite{K_Li1, K_Li2, K_Li3}, $^{161}$Dy-$^{40}$K\cite{Dy_K1, Dy_K2}, $^{173}$Yb-$^{6}$Li\cite{Yb_Li, Yb_Li2} and $^{53}$Cr-$^{6}$Li\cite{Cr_Li, Cr_Li2}. With mass-imbalance, it has been shown that the three-dimensional(3D) $(N+1)$ system, which consists of $N$ identical heavy fermions interacting with a light atom, can easily support cluster bound states. These bound states can be classified as universal and Efimov types. For instance,  the $(2+1)$ trimer emerging at the heavy-light mass ratio $\eta=8.2$\cite{KM} is a universal type, where the trimer energy does not rely on any microscopic detail of short-range potential but solely depends on the scattering length and the masses. In comparison, a sequence of Efimov-type trimers can emerge at $\eta=13.6$ with discrete scaling symmetry\cite{Efimov}, whose energies are non-universal as they additionally rely on a three-body parameter at short range. Similarly, the (3+1) tetramer\cite{Castin, Blume} and (4+1) pentamer\cite{Petrov}, of both universal and Efimov types, also exist in 3D above certain critical mass ratios. 
Unfortunately, these cluster bound states typically require a large mass imbalance, and none of them has been observed in laboratories till now.

In this context, the 2D cold atomic systems might be a more promising platform to achieve the goal. 
As inferred by the fact that any infinitesimal attraction in 2D can afford a two-body bound state, one expects an equally easier formation of cluster bound states in 2D fermion systems. Meanwhile, since there is no Efimov-type bound state in 2D,  it is a quite clean platform for the study of universal few- and many-body physics.   
However, right now the related knowledge on cluster bound states therein is quite limited. For the fermionic $(N+1)$ system, only the trimer formation at $\eta=3.34$\cite{Pricoupenko} and a tetramer at   
$\eta=5.0$\cite{Parish} were reported, both with total angular momentum $|m_{\rm tot}|=1$. 

In this work, we bring two new members to the family of  $(N+1)$ universal cluster bound states in 2D. They are the ground state tetramer ($N=3$) emerging at $\eta=3.38$ and the ground state pentamer ($N=4$) at $\eta=5.14$, both with total angular momentum $m_{\rm tot}=0$. 
The previously reported tetramer at $\eta>5.0$\cite{Parish} turns out to be an excited state instead. We have analyzed the inner structure of these cluster bound states and shown that they are associated with distinct angular momentum decompositions in the dimer-fermion frame. Because of this, they display very different momentum-space distributions and correlations for the heavy component. Interestingly, the ground state pentamer (and tetramer) features a spontaneous {\it triangular} crystallization with (without) center in the three-body (two-body) correlation function of heavy fermions. These universal bound states and their associated correlation patterns can be readily detected in a number of Fermi-Fermi mixtures now available in cold atoms, which further shed light on novel few-body correlations in the corresponding many-body systems.

We start from the following Hamiltonian ($\hbar=1$):
\begin{equation}
H=\sum_{\mathbf{k}} \left(\epsilon^a_{\mathbf{k}} a_{\mathbf{k}}^{\dagger} a_{\mathbf{k}} + \epsilon^f_{\mathbf{k}} f_{\mathbf{k}}^{\dagger} f_{\mathbf{k}}\right) +\frac{g}{S} \sum_{\mathbf{Q}, \mathbf{k}, \mathbf{k}^{\prime}} a_{\mathbf{Q}-\mathbf{k}}^{\dagger} f_{\mathbf{k}}^{\dagger} f_{\mathbf{k}^{\prime}} a_{\mathbf{Q}-\mathbf{k}^{\prime}}. \label{eq:H}
\end{equation}
Here $a_{\mathbf{k}}^{\dagger}$ and $f_{\mathbf{k}}^{\dagger}$ respectively create a light atom and a heavy fermion with momentum ${\cp k}$ and energy $\epsilon^{a,f}_{\mathbf{k}}=\mathbf{k}^2/(2m_{a,f})$ ($m_a<m_f$); the bare coupling $g$  is renormalized through $1/g=-1/S \sum_{\mathbf{k}}1/(\epsilon^a_{\mathbf{k}}+\epsilon^f_{\mathbf{k}}+E_{2b})$, where $S$ is the 2D area, and $E_{2b}=(2\mu a_{2D}^2)^{-1}$ is the two-body binding energy determined by scattering length $a_{2D}$ and reduced mass $\mu=m_am_f/(m_a+m_f)$. 

A general wave-function for the ($N+1$) bound state can be written in the center-of-mass (COM) frame as: 
\begin{equation}
|\Psi\rangle_{N+1} = \sum_{{\cp k}_1{\cp k}_2...{\cp k}_N} \psi_{{\cp k}_1{\cp k}_2...{\cp k}_N} a^{\dag}_{-{\cp k}_1-{\cp k}_2...-{\cp k}_N} f^{\dag}_{{\cp k}_1}f^{\dag}_{{\cp k}_2}...f^{\dag}_{{\cp k}_N}|0\rangle. \label{wf}
\end{equation}
Imposing the Schr{\"o}dinger equation $H|\Psi\rangle_{N+1}=E_{N+1}|\Psi\rangle_{N+1}$, one can obtain the integral equation for the function $F_{{\cp k}_2...{\cp k}_N}=\sum_{{\cp k}_1} \psi_{{\cp k}_1{\cp k}_2...{\cp k}_N}$, from which the binding energy $E_{N+1}$ can be extracted\cite{Pricoupenko2}:
\begin{align}
&F_{{\cp k}_2...{\cp k}_N}\left( \frac{S}{g} + \sum_{{\cp k}} \frac{1}{E_{{\cp k}{\cp k}_2...{\cp k}_N}}\right) = 
\nonumber\\
&\ \ \ \ \ \ \ \ \ \ \  \sum_{{\cp k}} \frac{\sum_{i=2}^N F_{{\cp k}_2...{\cp k}_{i-1}{\cp k}_i{\cp k}_{i+1}...{\cp k}_N}\delta_{{\cp k}{\cp k}_i}}{E_{{\cp k}{\cp k}_2...{\cp k}_N}}, \label{F}
\end{align}
with $E_{{\cp k}_1{\cp k}_2...{\cp k}_N}=-E_{N+1} +\epsilon^a_{-{\cp k}_1...-{\cp k}_N}+\sum_{i=1}^N\epsilon^f_{{\cp k}_i}$. The original coefficients in (\ref{wf}) can be related to the $F$-function via, for instance, $\psi_{{\cp k}_1{\cp k}_2}\propto (F_{{\cp k}_1}-F_{{\cp k}_2})/E_{{\cp k}_1{\cp k}_2}$ for $N=2$, $\psi_{{\cp k}_1{\cp k}_2{\cp k}_3}\propto (F_{{\cp k}_1{\cp k}_2}-F_{{\cp k}_1{\cp k}_3} +F_{{\cp k}_2{\cp k}_3})/E_{{\cp k}_1{\cp k}_2{\cp k}_3}$ for $N=3$, etc. One can see that both $\psi_{\{{\cp k}_i\}}$ and $F_{\{{\cp k}_i\}}$ are anti-symmetric with respect to the exchange ${\cp k}_i\leftrightarrow {\cp k}_j \ (i\neq j)$.

Physically, $F_{{\cp k}_2...{\cp k}_N}$ describes the relative motion between the heavy-light dimer (at momentum ${\cp k}\equiv-{\cp k}_2...-{\cp k}_N$) and the rest $N-1$ fermions (at ${\cp k}_2, ...,{\cp k}_N$). It can be factorized as
\begin{equation}
F_{{\cp k}_2...{\cp k}_N}=\sum_m \tilde{F}_m(k_2,...,k_N,\theta_{32},...\theta_{N2})e^{im\theta_2}, \label{factorized}
\end{equation}
where  ${\cp k}_i\equiv k_ie^{i\theta_i}$ and $\theta_{ij}=\theta_i-\theta_j$. It is straightforward to show that  $\{\tilde{F}_m\}$ are decoupled between different $m$, and thus in principle one can work with the reduced integral equation from (\ref{F}) for  $\tilde{F}_m$ in each $m$-sector. 
We note that the sector index $m$ actually represents the total angular momentum of the according $(N+1)$ system. This can be seen by expressing the total angular momentum operator as $\hat{L}_z=-i(\partial/\partial \theta+\sum_{j=2}^N \partial/\partial \theta_j)$, where $\theta$ is the angle of dimer momentum ${\cp k}$. Utilizing the identities $d/d\theta_j=\partial/\partial \theta_j+\partial/\partial \theta (d\theta/d\theta_j)$ and $\sum_{j=2}^N d\theta/d\theta_j=1$, we then have $\hat{L}_z= -i \sum_{j=2}^N d/d\theta_j$ and when act it on the wave-function (\ref{factorized}) in the $m$-sector we get its eigenvalue exactly as $m$. 
  


\begin{figure}[t]
\includegraphics[width=8.5cm]{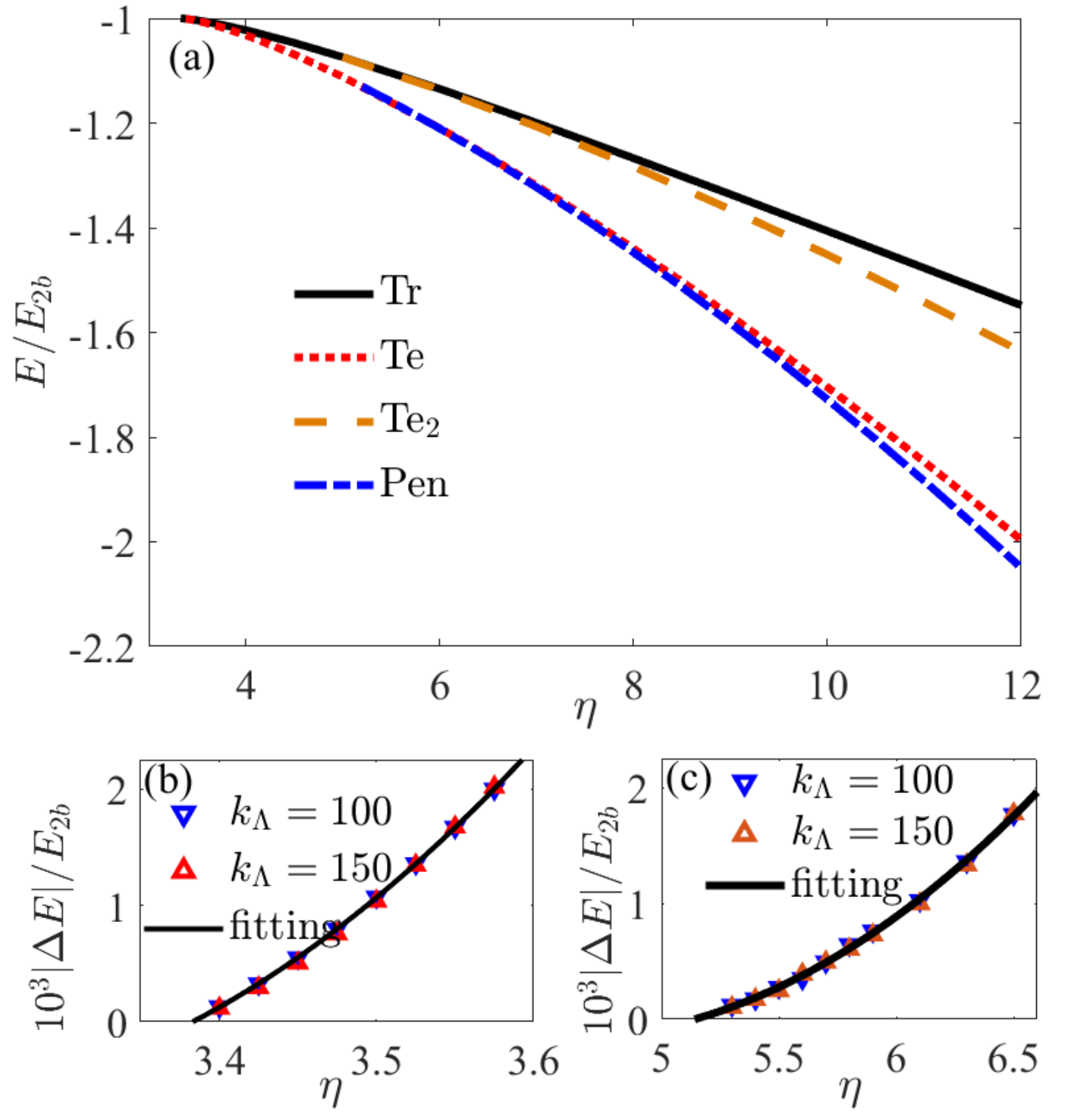}
\caption{(Color Online). (a) Energies of the ground state trimer ('Tr'), tetramer ('Te'), pentamer ('Pen'), and an excited tetramer ('Te$_2$') in 2D as functions of mass ratio $\eta\equiv m_f/m_a$. The energy unit $E_{2b}$ is the two-body binding energy. (b) Energy difference between the ground state tetramer('Te') and trimer('Tr'), $\Delta E=E_4-E_3$. (c)  Energy difference between the ground state pentamer('Pen') and tetramer('Te'), $\Delta E=E_5-E_4$. In (b,c), the triangle points are numerical results from different momentum cutoffs $k_{\Lambda}=100,\ 150$ (scaled by $1/a_{2D}$); the curves show binomial fittings according to $10^3|\Delta E|/E_{2b}=\alpha (\eta-\eta_c) + \beta (\eta-\eta_c)^2$, with parameters $(\alpha,\beta,\eta_c)=(7.08,\ 17.39, \ 3.38)$ (b), $(0.58,\ 0.52,\ 5.14)$ (c). }  \label{fig_E}
\end{figure}

To solve Eq.(\ref{F}), we have performed coarse graining of the module and the angle of $\{{\cp k}_j\}$ and transformed it into matrix equation. 
The anti-symmetry of $F$-function is carefully handled and the double counting of any ${\cp k}_i\leftrightarrow {\cp k}_j$ pair is avoided in setting up the matrix. 
For the trimer and tetramer, we directly solve the matrix equation to obtain the energies of ground state and excited states. For the pentamer case, given the large matrix size we resort to the iteration method to find the ground state. In all cases,  the resulted $F$-functions are checked to perfectly match the form (\ref{factorized}) with certain $m=m_{\rm tot}$.
In our numerics, the convergence of energy is further guaranteed by choosing different discretization schemes and different momentum cutoff $k_{\Lambda}$. 

The results of $E_{N+1}$ for $N=2$(trimer), $3$(tetramer) and $4$(pentamer) are shown in Fig.\ref{fig_E}(a) as functions of mass ratio $\eta\equiv m_f/m_a$. We first reproduce the lowest trimer ('Tr') as studied in Ref.\cite{Pricoupenko}, which emerges from the dimer threshold at $\eta=3.34$ and is double degenerate with $m_{\rm tot}=\pm 1$. We then turn to the tetramer and pentamer solutions. In contrast to the trimer case, we find the lowest tetramer ('Te') and the lowest pentamer ('Pen') are both non-degenerate and associated with $m_{\rm tot}=0$. The previously reported tetramer in $m_{\rm tot}=\pm 1$ sector\cite{Parish}, denoted as 'Te$_2$' in Fig.\ref{fig_E}(a), is an excited state emerging at a larger $\eta$ and with a higher energy than 'Te'.  To pin down the critical $\eta_c$ for the emergence of the lowest tetramer and pentamer, we have used the binomial fitting of the energy difference near $\eta\sim \eta_c$, i.e., $10^3|E_{N+1}-E_{N}|/E_{2b}=\alpha (\eta-\eta_c) + \beta (\eta-\eta_c)^2$, with $\alpha,\beta,\eta_c$ all fitting parameters. As shown in Fig.\ref{fig_E}(b,c), the fittings give the critical mass ratio for the lowest tetramer (below trimer) as $\eta_c=3.38$, and for the lowest pentamer (below tetramer) as  $\eta_c=5.14$. 
We emphasize that all these bound states are  {\it universal}, in that their binding energies do not depend on the momentum cutoff $k_{\Lambda}$ (or short-range details), as can be seen from Fig.\ref{fig_E}(b,c).


To understand the emergence of these cluster states associated with different $m_{\rm tot}$, it is  useful to decompose  the total angular momentum
 in the dimer-fermion relative motion frame.  Here we introduce a new set of coordinates $\bar{\cal K}\equiv ({\cp K}, \bar{\cp k}_2, ..., \bar{\cp k}_N)^T$, where ${\cp K}={\cp k}+\sum_{j=2}^N{\cp k}_j$ is the COM momentum and $\bar{\cp k}_j=[(m_a+m_f){\cp k}_j-m_f{\cp k}]/(2m_f+m_a)$ is the relative momentum between the dimer at ${\cp k}$ and the fermion at ${\cp k}_j$. In this way, $\bar{\cal K}$ can be related to the original coordinate vector ${\cal K}\equiv ({\cp k}, {\cp k}_2, ..., {\cp k}_N)^T$ via $\bar{\cal K}=A{\cal K}$, with $A$ the transformation matrix. It is then straightforward to prove that the total angular momentum operator, $\hat{L}_z=(-i/2){\cal K}^T\times \partial/\partial {\cal K}$, can be re-expressed as $\hat{L}_z=(-i/2)\bar{\cal K}^T\times \partial/\partial \bar{\cal K}$. In the COM frame (${\cp K}=0$), we  have
 \begin{align}
&\ \ \ \ \ \  \ \ \ \ \hat{L}_z= \sum_{j=2}^{N} \hat{l}_{z,j}, 
& \ \ {\rm with} \ \  \hat{l}_{z,j}= -\frac{i}{2} \bar{\cp k}_j \times \frac{\partial}{\partial \bar{\cp k}_j} \label{Ll}
\end{align}
which tells that the total angular momentum ($m_{\rm tot}$) is just the sum of all relative angular momenta ($\{\bar{m}_j$\}) between the  heavy-light dimer and the rest fermions, i.e. $m_{\rm tot}=\sum_{j=2}^N \bar{m}_j$.

\begin{figure}[t]
\includegraphics[width=8.5cm]{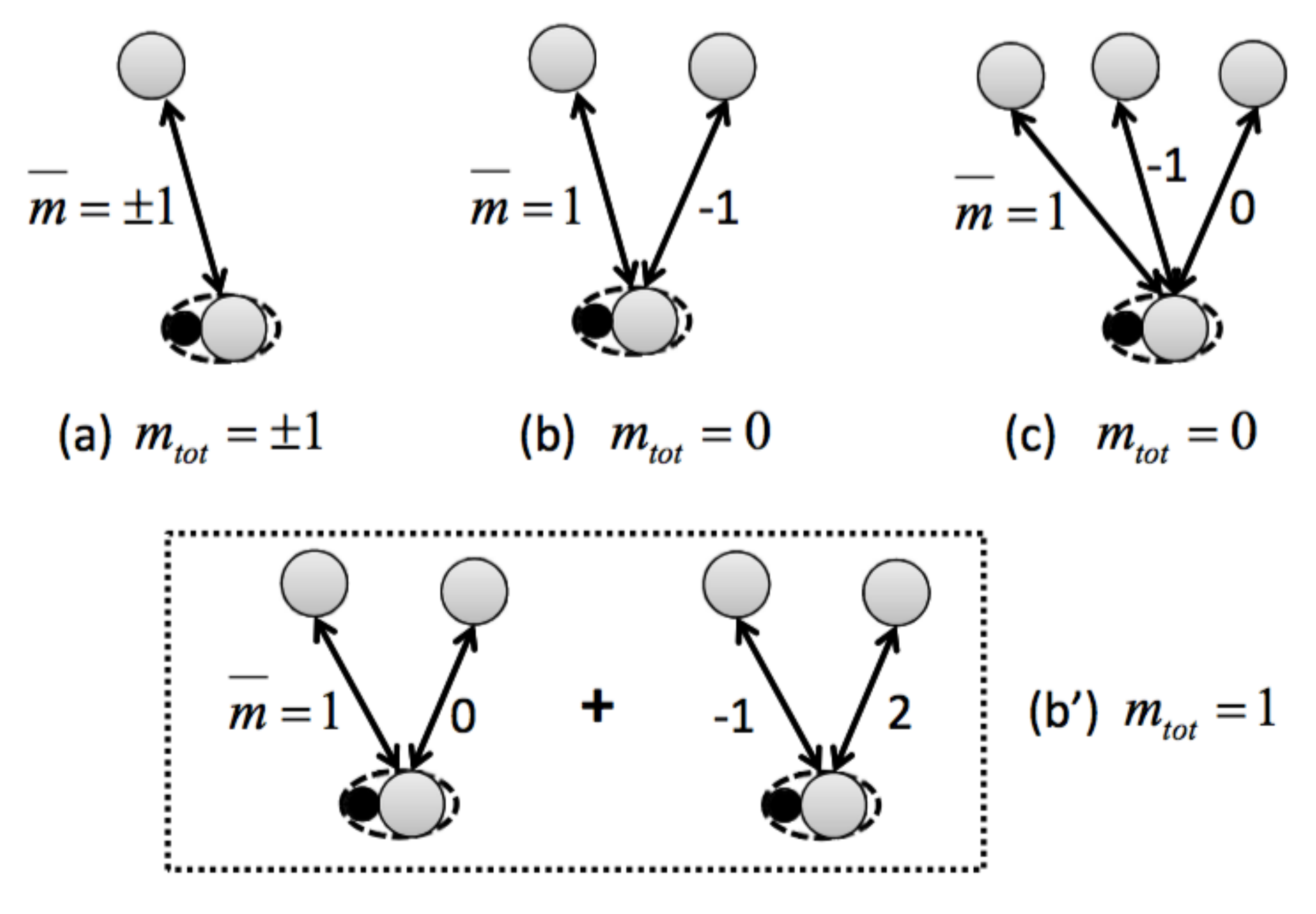}
\caption{(Color Online). Illustration of angular momentum decomposition in the dimer-fermion relative motion frame. For the ground state trimer (a), the relative angular momentum between the dimer and the fermion is $\bar{m}=\pm 1$, which is also the total angular momentum $m_{\rm tot}$. For the ground state tetramer (b), the two relative angular momenta are $\bar{m}=1$ and $-1$, giving $m_{\rm tot}=0$. For the ground state pentamer, the three relative angular momenta are dominated by $\bar{m}=1, \ -1,\ 0$, again giving $m_{\rm tot}=0$. In (b'), we show two possible decomposition configurations for the excited tetramer with $m_{\rm tot}=1$, either with $\bar{m}=1,0$ or $-1,2$. }  \label{fig_illustration}
\end{figure}

Eq.(\ref{Ll}) provides a powerful tool for analyzing the physical origin of various cluster bound states, as illustrated in Fig.\ref{fig_illustration}. We start from the ground state trimer in Fig.\ref{fig_illustration}(a), where there is only one dimer-fermion pair and therefore its relative angular momentum is exactly the total angular momentum of the system, i.e., $\bar{m}=m_{\rm tot}=\pm 1$.  Given such double degeneracy of the lowest dimer-fermion channel, it is then natural to expect that in the ground state tetramer, the two relative angular momenta between the dimer and two fermions  are respectively $\bar{m}=1$ and $-1$, thus giving $m_{\rm tot}=0$ (Fig.\ref{fig_illustration}(b)). This combination implies that once the trimer appears with $\bar{m}=\pm 1$, the tetramer can also be easily supported --- this explains why the  critical mass ratio for the trimer formation ($=3.34$) is so close to the tetramer one ($=3.38$). Further,  for the ground state pentamer, the three dimer-fermion angular momenta are dominated by $\bar{m}=1,-1$ and $0$, again leading to $m_{\rm tot}=0$ (Fig.\ref{fig_illustration}(c)). In Fig.\ref{fig_illustration}(b'), we show the angular momentum decomposition for the excited tetramer with $m_{\rm tot}=1$, where the two dimer-fermion lines are either with $\bar{m}=1,0$ or $-1,2$. Clearly these combinations are less energetically favored as compared to $\bar{m}=1,-1$ and thus they should stay in an excited manifold. 


\begin{figure}[t]
\includegraphics[width=8cm]{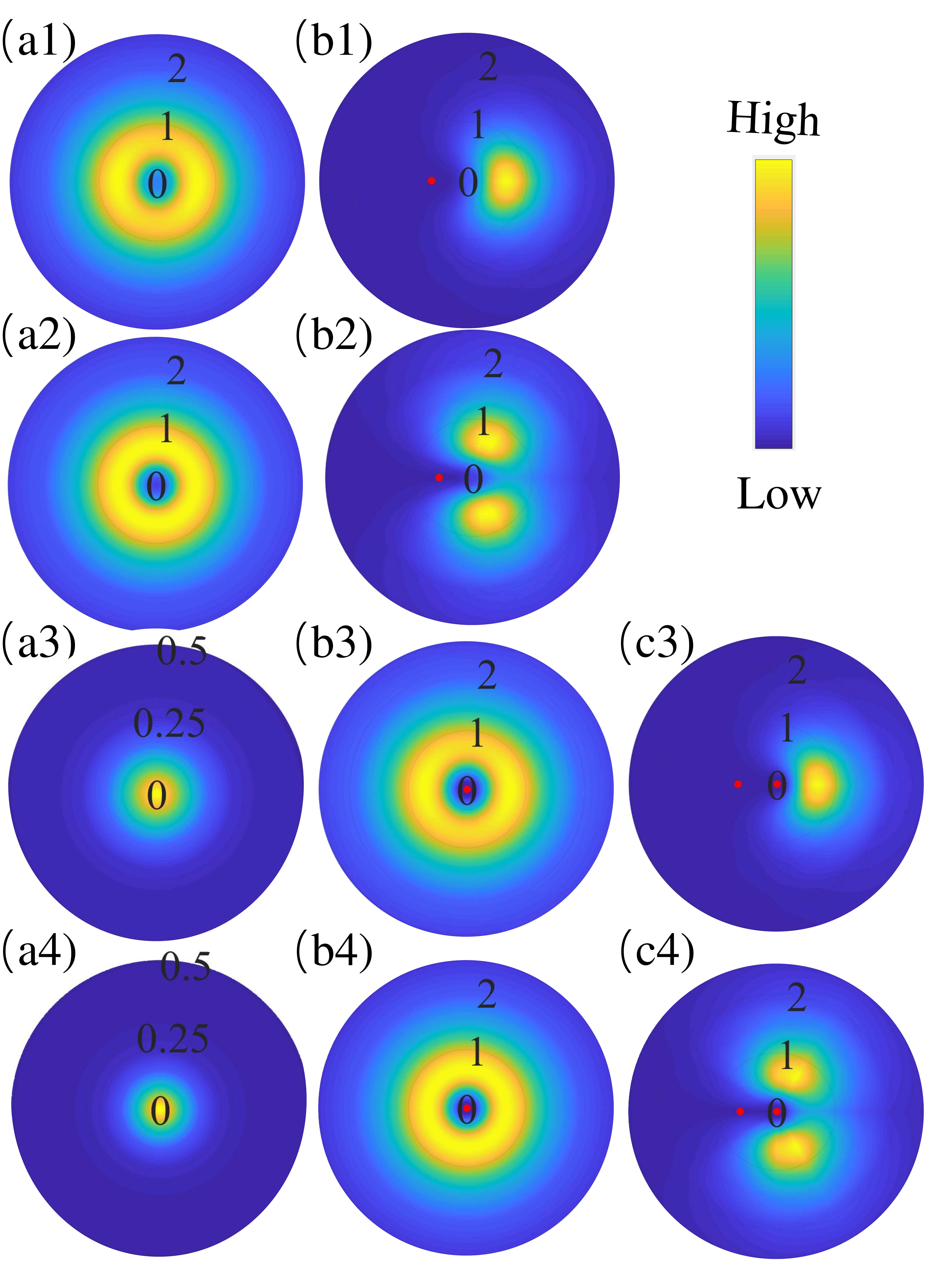}
\caption{(Color Online). Momentum(${\cp k}$)-space correlations of heavy fermions for the lowest trimer (a1,b1), lowest tetramer (a2,b2,c2), excited tetramer (a3,b3,c3) and lowest pentamer (a4,b4,c4). (a1-a4) show the one-body density distribution $G^{(1)}({\cp k})$; (b1-b4) show the two-body correlation $G^{(2)}({\cp k}_1,{\cp k})$, with ${\cp k}_1$  pinned down at the maximum of $G^{(1)}$ (red point); (c3-c4) show the three-body correlation $G^{(3)}({\cp k}_1,{\cp k}_2,{\cp k})$, with ${\cp k}_1$ and ${\cp k}_2$ pinned down, respectively, at the maximum of $G^{(1)}$ and $G^{(2)}$ (red points). 
Here we take the mass ratio $\eta=40/6$, and the momentum unit is $1/a_{2D}$.  }  \label{fig_correlation}
\end{figure}

The distinct inner structures of these bound states can be reflected in their momentum-space correlations. To see this, we examine the $n$-body density correlation function of heavy fermions, as defined by
\begin{equation}
G^{(n)}({\cp k}_1,{\cp k}_2,...,{\cp k}_n)=\langle n^f_{{\cp k}_1} n^f_{{\cp k}_2} ...n^f_{{\cp k}_n}\rangle,
\end{equation}
with $n^f_{\cp k}$ the density operator of fermion at momentum ${\cp k}$. For $n=1$, $G^{(1)}({\cp k})=\langle n^f_{\cp k}\rangle$ is exactly the one-body density distribution. For higher $n(\ge 2)$, $G^{(n)}$ gives the density correlation pattern of $n$ fermions within the system. In Fig.\ref{fig_correlation} we take $\eta=40/6$ and show the typical pattern of $G^{(n)}$ in ${\cp k}$ space, with $n$ up to $3$, for various cluster bound states. Our strategy in plotting $G^{(2)}$ and $G^{(3)}$ is as follows. For $G^{(2)}({\cp k}_1,{\cp k})$, we have fixed  ${\cp k}_1$  at the maximum of one-body density distribution $G^{(1)}$, and for $G^{(3)}({\cp k}_1,{\cp k}_2,{\cp k})$ we take the same strategy of ${\cp k}_1$ and further fix ${\cp k}_2$ at the maximum of $G^{(2)}({\cp k}_1,{\cp k}_2)$. Apparently such strategy is able to extract the most dominant correlation pattern of heavy fermions.  


From Fig.\ref{fig_correlation} (a1) and (a2), we can see that the ground state trimer and tetramer, although associated with different $m_{\rm tot}$, have quite similar one-body density distribution in ${\cp k}$ space, which is peaked on a circle with finite radius (determined by the binding energy). This can be attributed to the fact that their angular momentum decompositions in the dimer-fermion frame are all composed by $\bar{m}=\pm 1$ (see Fig.\ref{fig_illustration} (a,b)), which ensure zero weight at the origin (${\cp k}=0$) and the largest weight at finite $|{\cp k}|$. However, since the tetramer has two dimer-fermion pairs while the trimer only has one, their corresponding two-body correlations are very different. As shown in Fig.\ref{fig_correlation}(b1) and (b2), the trimer shows a clear diagonal correlation and the tetramer shows a stable triangular correlation. 

In comparison, the excited tetramer and the lowest pentamer exhibit a distinct type of distribution/correlation patterns. Because of the  presence of $\bar{m}=0$ component in their angular momentum decomposition (Fig.\ref{fig_illustration} (c,b')), their one-body density distributions are both peaked at ${\cp k}=0$, see Fig.\ref{fig_correlation} (a3) and (a4). The manifestation of $\bar{m}=\pm 1$ components only show up in the level of two- and three-body correlations. For instance, their two-body correlations are both peaked at a finite radius (Fig.\ref{fig_correlation} (b3,b4)) and the three-body correlations show crystalline patterns, either as a centered diagonal  (see Fig.\ref{fig_correlation} (c3) for the excited tetramer) or as a centered triangle (Fig.\ref{fig_correlation} (c4) for the lowest pentamer).  

We remark that the emergent crystalline patterns, as shown in Fig.\ref{fig_correlation} (b1,b2,c3,c4),  exactly represent the dominant distributions of $N$ correlated fermions in the ($N+1$) cluster bound states. In particular, the relative distribution of three fermions in the lowest tetramer state automatically form a regular triangle (Fig.\ref{fig_correlation}(b2)), and the four fermions in the lowest pentamer form a centered triangle (Fig.\ref{fig_correlation} (c4)). In distinct contrast to the Pauli crystal for identical fermions confined in a harmonic trap\cite{Pauli1, Pauli2, Jochim_expt1, Pauli3_Jason}, here the  crystalline patterns are purely driven by the heavy-light interaction and thus may be termed as an interaction-induced crystal. 

In cold atoms experiment, the one-body density distribution in momentum space can be directly measured through the time of flight technique, and the two-body density correlation can be measured using the atom noise in absorption images\cite{theory_noise, expt_noise1,expt_noise2,expt_noise3,expt_noise4,expt_noise5}  or the single atom resolved image\cite{Jochim_expt2}. Using the latter technique,  the three- and even higher-body correlations now also become measurable as in the recent observation of Pauli crystals\cite{Jochim_expt1}. 

To summarize, we have reported the ground state tetramer and pentamer formation in the 2D $(N+1)$ cluster system, as long as the mass ratios between the heavy fermions and the light atom are, respectively, beyond $3.38$ and $5.14$. These bound states require considerably lower mass imbalance than the 3D counterparts, and thus can now be accessible by more Fermi-Fermi mixtures such as $^{40}$K-$^{6}$Li\cite{K_Li1, K_Li2, K_Li3}, and $^{53}$Cr-$^{6}$Li\cite{Cr_Li, Cr_Li2} (the tetramer is also accessible by $^{161}$Dy-$^{40}$K\cite{Dy_K1, Dy_K2}). These ground state tetramer and pentamer  both emerge in $m_{\rm tot}=0$ channel. Nevertheless, they can be distinguished by the momentum-space distribution/correlation patterns of heavy fermions (Fig.\ref{fig_correlation}), thanks to their distinct angular momentum decompositions in the dimer-fermion frame(Fig.\ref{fig_illustration}).

The universal feature of these bound states ensures that there is only one length scale, the 2D scattering length ($a_{2D}$), to determine their characteristic sizes. In the regime where $a_{2D}$ is much larger than the interaction range $r_0$, these bound states have very little overlap with deep molecules (of size $\sim r_0$) and therefore should be quite stable against inelastic loss\cite{inelastic_2d_1,inelastic_2d_2}. 
Moreover, our theory is expected to apply for a realistic quasi-2D setup under strong axial trap, i.e., when the trap length $l_z\ll a_{2D}$. In future, it is worthwhile to explore how these 2D bound states connect to the 3D counterparts\cite{Castin, Blume, Petrov} by changing the confinement strength, as previously studied for small clusters\cite{Parish,Parish2}.

Our results shed light on novel few-body correlations in mass-imbalanced Fermi-Fermi mixtures. Specifically, by adjusting the mass ratio between fermion components,  various cluster bound states can be supported and they strongly suggest the dominant multi-body correlations well beyond the two-body one. 
As an example, we have shown recently that these bound states can be dressed with particle-hole excitations in Fermi polarons and drive a sequence of smooth crossover therein\cite{Liu}, in distinct contrast to the equal mass case. For a thermodynamic system, these bound states may induce a new fermion superfluid based on cluster condensation, which is far beyond the conventional pairing mechanism. We hope our current work can stimulate more studies in future on such intriguing phases in mass-imbalanced fermions.  

\bigskip

{\it Acknowledgements.} 
The work is supported by the National Key Research and Development Program of China (2018YFA0307600), the National Natural Science Foundation of China (12074419, 12134015), and the Strategic Priority Research Program of Chinese Academy of Sciences (XDB33000000).

\end{document}